\documentclass[twocolumn,secnumarabic,amssymb, nobibnotes, aps, prc, showpacs]{revtex4-1}

\newcommand{\abs}[1]{\vert{#1}\vert}
\usepackage{graphicx,amsmath}
\usepackage{color}

\begin{document}

\title{Thouless--Valatin Rotational Moment of Inertia from the Linear Response Theory}%

\author{Kristian Petr\'ik}%
\email[]{kristian.k.petrik@jyu.fi}
\affiliation{Helsinki Institute of Physics, University of Helsinki, P.O. Box 64, FI-40014 Helsinki, Finland}
\affiliation{Department of Physics, University of Jyvaskyla, P.O. Box 35, FI-40014 Jyvaskyla, Finland}
\author{Markus Kortelainen}%
\email[]{markus.kortelainen@jyu.fi}
\affiliation{Department of Physics, University of Jyvaskyla, P.O. Box 35, FI-40014 Jyvaskyla, Finland}
\affiliation{Helsinki Institute of Physics, University of Helsinki, P.O. Box 64, FI-40014 Helsinki, Finland}
\date{\today}%
\begin{abstract}
Spontaneous breaking of continuous symmetries of a nuclear many--body system results in appearance of zero--energy restoration modes. Such modes introduce a non--physical contributions to the physical excitations called spurious Nambu--Goldstone modes. Since they represent a special case of collective motion, they are sources of important information about the Thouless--Valatin inertia. The main purpose of this work is to study the Thouless--Valatin rotational moment of inertia as extracted from the Nambu--Goldstone restoration mode that results from the zero--frequency response to the total angular momentum operator. We examine the role and effects of the pairing correlations on the rotational characteristics of heavy deformed nuclei in order to extend our understanding of superfluidity in general. We use the finite amplitude method of the quasiparticle random phase approximation on top of the Skyrme energy density functional framework with the Hartree--Fock--Bogoliubov theory. We have successfully extended this formalism and established a practical method for extracting the Thouless--Valatin rotational moment of inertia from the strength function calculated in the symmetry restoration regime. Our results reveal the relation between the pairing correlations and the moment of inertia of axially deformed nuclei of rare--earth and actinide regions of the nuclear chart.      
We have also demonstrated the feasibility of the method for obtaining the moment of inertia for collective Hamiltonian models. We conclude that from the numerical and theoretical perspective, the finite amplitude method can be widely used to effectively study rotational properties of deformed nuclei within modern density functional approaches.     
\end{abstract}


\maketitle

\section{Introduction}

Collective modes or excitations of many--fermion systems are excellent sources of vital information about the properties of the effective interaction among the constituent particles and correlations that govern the dynamics of the many--body system. It is a major objective for the nuclear theory to extensively describe all types of collective nuclear motion. Studying the low--lying excitations brings better understanding of the properties of pairing correlations, shell structure or nuclear deformation, while the higher--lying giant nuclear resonances reveal essential details about nuclear photoabsorption reactions or nuclear matter symmetry energy and compressibility \cite{Ring80, Harakeh01, Arnould07, Ebata15}. One can analyze a response of a nucleus to a time--dependent external field to gain important knowledge about such excited states. 

There exist several groups of models dealing with the many--body structure and low--energy dynamics of atomic nucleus, for example interacting shell--model approaches, {\it ab--initio} methods, or the 
density functional theory (DFT) formalism. 
The nuclear DFT, belonging to the family of the self--consistent mean--field (SCMF) approaches, determines the nuclear one--body mean--field by using 
an energy density functional (EDF) adjusted for a given task~\cite{Bender03}. Withing the SCMF approach, the nuclear mean--field
is obtained by solving Hartree--Fock (HF) or Hartree--Fock--Bogoliubov (HFB) equations self--consistently. Currently, the nuclear DFT can rather successfully
describe the nuclear ground state properties throughout the whole nuclear chart~\cite{Erler12,Afa13,Kortelainen14}.

An important ingredient of the nuclear DFT approach is the concept of a spontaneous symmetry breaking. A mean--field wave function, obtained self--consistently, usually
breaks some of the symmetries of the full Hamiltonian. In principle, an exact ground state wave function of a nucleus does not break symmetries 
of the underlying many--body Hamiltonian, however, such kind of wave--function is often computationally out of reach and simplifying approximations have to be used. 
Effectively, the spontaneous symmetry breaking allows to introduce diverse short--range and long--range correlations to the deformed wave function.

In order to access various excitation modes, one needs to go beyond the static mean--field. One of the most utilized methods is the \textit{linear response theory}, that is,
the random--phase approximation (RPA) theory. Excellent progress has been achieved with the application of the RPA and the quasiparticle 
random--phase approximation (QRPA), the superfluid 
extension of the RPA, to the excited collective states of nuclei within the nuclear DFT framework.
Modern approaches usually employ the self--consistent RPA or QRPA calculations together 
with well--established EDFs~\cite{Bender03,Terasaki10,Terasaki11,Peru11,Martini11}. 
However, traditional approaches, like the very successful matrix formulation of the quasiparticle random--phase approximation (MQRPA), suffer from the fact that they are computationally heavy, especially when spherical symmetry becomes broken. This is the main reason why the QRPA formalism has been able to treat the deformed nuclei only recently.      

To deal with various drawbacks of standard methods for collective excitations, a very efficient formalism has been proposed. The so--called \textit{finite amplitude method} (FAM) was introduced, first to calculate the RPA strength functions \cite{Nakatsukasa07} and later extended to spherically symmetric QRPA (FAM--QRPA) \cite{Avogadro11}. The FAM--QRPA promptly became a powerful tool to effectively handle a broad range of nuclear phenomena. For example, it was successfully applied to study deformed axially symmetric nuclear systems within the HFB--Skyrme framework, individual QRPA modes, beta decay modes, collective moments of inertia, quadrupole and octupole strengths or the giant dipole resonance (GDR)~\cite{Stoitsov11, Hinohara13, Niksic13, Mus14, Hinohara15, Kortelainen15, Oishi16, Wang17}. Due to many distinct advantages, we chose the FAM--QRPA as the main theoretical framework for the purposes of this work.

The properties and dynamics
of heavy nuclear systems can provide helpful insights for the physics of confined atomic fermions, especially when 
superfluid characteristics are of interest~\cite{Farine2000, Urban2003, Urban2005, Zwierlein2005, Bausmerth2008, Riedl2011, Bulgac2013}. 
Although the length and energy scales of confined atomic systems are quite different to those of nuclear scale, the rotating superfluid Fermi liquid 
drop picture of a nucleus covers the essential concepts to build up a connection between these two fields. 
One of the connecting ingredients is the moment of inertia, which is believed to provide an unambiguous signature of superfluidity in general. Since the similarities between nuclear systems and trapped fermions are strong, studying the rotational moment of inertia of heavy deformed nuclei by the best available theoretical methods will offer crucial information about the phenomenon of superfluidity. This paper is thus aimed to present results that will be valuable not only for the nuclear physics, but for the physics of trapped fermionic gases as well. Such universality can also help to study the transition between macroscopic and microscopic fermionic systems.    

It is well know that the experimental moments of inertia of nuclei are usually notably smaller than the corresponding rigid body values~\cite{BM75}. When deriving the expression for the moment of inertia, the simplest case leads to the Inglis moment of inertia, which represents the response of free gas. The moment of inertia resulting from this basic Inglis formula is typically very close to the rigid body values. In order to improve the description of the data, one has to take into account the pairing correlations, which lead, within the BCS formalism, to the so--called Belyaev formula. This superfluid expansion of the Inglis inertia lowers the theoretical values towards the experimental results, which manifests the importance of the correlations of the pairing type. 

More generally, one can obtain the moment of inertia from the self--consistent cranking theory or the linear response theory as derived by Thouless and Valatin~\cite{Thouless62}. In this context, the inertia takes into account also the nuclear response, i.e. the effects of induced fields that represent a reaction of the system to an external perturbation, determined by a given operator. In this way, the moment of inertia is defined by a generalized expression that contains, as special cases, both Inglis and Belyaev formulas.  

Although the moment of inertia of superfluid nuclei is significantly smaller than the rigid body inertia, it is still larger than that of a strictly irrotational motion. From this observation, we can conclude that the currents in superfluid rotating nuclei have a two--component character; the total current consists of rotational and irrotational components and the moment of inertia is the key observable to determine the effects of pairing correlations on the collective motion of nuclear systems. Analogous behavior is found to be present in the trapped superfluid Fermi gases at zero and non--zero temperatures, where the temperature dependence is shown to have crucial effects on the moment of inertia as well as the qualitative behavior of the currents~\cite{Urban2003,Urban2005}.  

The main objective of this work is to apply the FAM--QRPA approach to study the Thouless--Valatin (TV) rotational moment of inertia in various nuclear systems 
and inspect the significance of the pairing correlations with respect to the rotational characteristics of studied nuclei. The TV inertia can be obtained 
from the spurious zero--frequency (zero--energy) mode, which is a consequence of a broken symmetry. We propose an efficient and numerically accessible method for extracting this quantity from the total angular momentum operator that acts in a role of an external perturbation. This method can be easily used to describe collective modes of a wide variety of deformed nuclei.  

As a proof of concept, we will also demonstrate the usefulness of our method in providing local QRPA calculations, specifically constrained calculations of the rotational moment of inertia as a function of density. In this way, FAM--QRPA could be used to easily access collective mass parameters that represent a key input for microscopic collective Hamiltonian models.       

Our article is organized in the following way. In Sec. II, an overview of the theoretical framework is presented, in which the time--dependent HFB theory and FAM--QRPA approach are discussed in detail. Additionally, we briefly analyze the spontaneous symmetry breaking phenomenon and its connection to the collective TV inertia. In Sec. III, the numerical setup and main FAM--QRPA results are covered. Final conclusions and prospects are given in Sec. IV.   
  
\section{Theoretical framework}
In this section we recapitulate all necessary theoretical details and go through the essential expressions that were used as a foundation for our calculations. Main ideas and assets of the FAM--QRPA will be discussed. 

\subsection{Finite amplitude method}
A static mean--field approach without pairing correlations (Hartree--Fock) or with pairing correlations (Hartree--Fock--Bogoliubov) can basically provide the nuclear binding energy 
and other ground--state bulk properties. In order to access excited states and dynamics of the system, one has to go beyond the mean--field formalism.
One approach is to apply time--dependent extensions to the stationary mean--field models. 
In the following, we will use the quasiparticle random--phase approximation as a theoretical method that represents a small--amplitude limit of
the time--dependent superfluid HFB (TD--HFB) theory.

The HFB ground state $|\Phi\rangle$ is obtained by the minimization of the total energy defined through the energy density 
$\mathcal{E}(\rho,\kappa,\kappa^*)$. 
In general, expectation values of an operator with an HFB state can be expressed with one--body densities. 
In the quasiparticle picture, the energy density $\mathcal{E}(\rho,\kappa,\kappa^*)$ introduces the one--body 
density matrix $\rho$ and the pairing tensor $\kappa$,
\begin{align}
\rho_{ij} &=\langle\Phi|\hat c^+_j \hat c_i|\Phi\rangle =(V^\ast V^T)_{ij} = \rho^\ast_{ji}, 
\nonumber\\
\kappa_{ij} &=\langle\Phi|\hat c_j \hat c_i|\Phi\rangle =(V^\ast U^T)_{ij} = -\kappa_{ji},
\end{align} 
where $V$ and $U$ are the Bogoliubov transformation matrices that define  the linear relation between the Bogoliubov quasiparticles $\hat a_\mu^+,\hat a_\mu$ and the bare particles $\hat c_k^+,\hat c_k$,
\begin{align}
\hat a_\mu^+ &=\sum_k \left[U_{k\mu}\hat c_k^+ + V_{k\mu}\hat c_k \right] ,
\nonumber\\
\hat a_\mu &=\sum_k \left[U_{k\mu}^\ast\hat c_k^+ + V_{k\mu}^\ast\hat c_k \right].
\end{align}

The single--particle Hamiltonian $h$ and the pairing potential $\Delta$ follow from the variation of the energy density functional $\mathcal{E}$ with respect to $\rho$ and $\kappa^\ast$,
\begin{align}
h_{ij}(\rho,\kappa,\kappa^\ast)=\frac{\partial \mathcal{E}}{\partial \rho_{ij}}, \hspace{20pt}
\Delta_{ij}(\rho,\kappa,\kappa^\ast)=\frac{\partial \mathcal{E}}{\partial \kappa^\ast_{ij}}.
\end{align}

By minimizing the total Routhian, one can derive the HFB equations in terms of the transformation matrices $V$ and $U$,
\begin{align}
\begin{pmatrix}
  h -\lambda   & \Delta     \\
  -\Delta^\ast & -h^\ast + \lambda 
\end{pmatrix}
\begin{pmatrix}
  U_\mu \\
  V_\mu 
 \end{pmatrix}
 =E_\mu
 \begin{pmatrix}
  U_\mu \\
  V_\mu
 \end{pmatrix},
\end{align}
where $E_\mu$ are the quasiparticle energies and $\lambda$ is the chemical potential introduced in order to fix the average particle number. 

The TD--HFB can be used to describe the time evolution of the quasiparticles under a one--body external perturbation that induces a polarization on the HFB ground state. Such perturbation can be expressed as a time--dependent field in a conjugated form,
\begin{align}
\hat F(t)=\eta\left[\hat F(\omega)e^{-i\omega t}+\hat F^\dagger(\omega)e^{i\omega t}\right] ,
\label{eq:Ft}
\end{align}
where $\eta$ is a small real parameter introduced for the purpose of the small amplitude approximation, used in the FAM--QRPA.

The time evolution of a quasiparticle operator under the influence of the external field that forces the oscillations can 
be expressed as the TD--HFB equation, 
\begin{align}
i\hbar\frac{\partial}{\partial t} \hat a_\mu(t)=\left[\hat H(t) + \hat F(t), \hat a_\mu(t)\right] ,
\end{align}
where the quasiparticle oscillations are given as
\begin{align}
\hat a_\mu(t)&=\left[\hat a_\mu + \delta\hat a_\mu(t)\right]e^{i E_\mu t} ,
\nonumber\\
\delta\hat a_\mu(t)&=\eta\sum_\nu \hat a_\nu^+ \left[X_{\nu\mu}(\omega)e^{-i\omega t}+Y_{\nu\mu}^\ast(\omega) e^{i\omega t}\right] ,
\end{align}
where $X_{\nu\mu}$ and $Y_{\nu\mu}$ are the FAM--QRPA amplitudes and $E_\mu$ is the one--quasiparticle energy.

The time--independent part $\hat F(\omega)$ of the one--body external perturbation $\hat F(t)$ can be now expressed (under the linear approximation) in the quasiparticle space as  
\begin{align}
\hat F(\omega) = \sum_{\mu<\nu}\left[F_{\mu\nu}^{20}(\omega)\hat a_\mu^+ \hat a_\nu^+ + \hat F_{\mu\nu}^{02}(\omega)\hat a_\mu \hat a_\nu\right] \,.
\label{eq:Fomega}
\end{align}

The external perturbation induces oscillations of the density atop the static HFB solution, thus the self--consistent Hamiltonian will contain also the induced part
as
$\hat H(t)=\hat H_{\text{HFB}} +\delta \hat H(t)$,
where the oscillating part $\delta \hat H(t)$ is defined similarly as the external field in Eq.~(\ref{eq:Ft}).

The response of the self--consistent Hamiltonian, given by the induced time--independent matrices $\delta H_{\mu\nu}^{20}$ and $\delta H_{\mu\nu}^{02}$, 
can be expressed with the HFB matrices $U$ and $V$ and induced fields $\delta h$, $\delta \Delta$ and $\overline{\delta\Delta}$ as
\begin{align}
\delta H_{\mu\nu}^{20}(\omega)=&\big[U^{\dagger}\delta h (\omega) V^\ast - V^\dagger \delta h(\omega)^T U^\ast 
\nonumber\\
&-V^\dagger\overline{\delta\Delta}(\omega)^\ast V^\ast+U^\dagger\delta\Delta(\omega)U^\ast\big]_{\mu\nu},
\nonumber\\
\delta H_{\mu\nu}^{02}(\omega)=&\big[U^{T}\delta h (\omega)^T V - V^T \delta h(\omega) U 
\nonumber\\
&-V^T \delta\Delta(\omega) V+U^T \overline{\delta\Delta}(\omega)^\ast U\big]_{\mu\nu}.
\end{align}

Here, the fields $\delta h$, $\delta \Delta$ and $\overline{\delta\Delta}$ were obtained through an explicit linearization of the Hamiltonian and can be thus expressed through fields linearized with respect to perturbed densities, namely $h^\prime$ and $\Delta^\prime$,
\begin{align}
\delta h(\omega) &= h^\prime \left[\rho_{\rm f},\kappa_{\rm f},\bar \kappa_{\rm f} \right],
\nonumber\\
\delta \Delta(\omega) &= \Delta^\prime \left[\rho_{\rm f},\kappa_{\rm f} \right],
\nonumber\\
\overline{\delta \Delta}(\omega)&=\Delta^\prime \left[\bar\rho_{\rm f},\bar\kappa_{\rm f} \right],
\end{align}
where the non--Hermitian density matrices depend on the external perturbation through the FAM--QRPA amplitudes, 
\begin{align}
\rho_{\rm f}(\omega)&= +U X(\omega)V^T + V^\ast Y(\omega)^T U^\dagger,
\nonumber\\
\bar \rho_{\rm f}(\omega) &=+ V^\ast X(\omega)^\dagger U^\dagger + U Y(\omega)^\ast V^T,
\nonumber\\
\kappa_{\rm f}(\omega)&= - U X(\omega)^T U^T - V^\ast Y(\omega)V^\dagger, 
\nonumber\\
\bar\kappa_{\rm f}(\omega) &= - V^\ast X(\omega)^\ast V^\dagger - U Y(\omega)^\dagger U^T. 
\end{align}

To access a transition strength function one needs to obtain the FAM--QRPA amplitudes $X_{\mu\nu}(\omega)$ and $Y_{\mu\nu}(\omega)$.
By using the linear approximation (i.e. linear response) one can derive the FAM--QRPA equations, given as
\begin{equation}
\begin{aligned}
X_{\mu\nu}(\omega)&=-\frac{\delta H_{\mu\nu}^{20}(\omega)-F_{\mu\nu}^{20}(\omega)}{E_\mu + E_\nu-\omega} \,,
\\
Y_{\mu\nu}(\omega)&=-\frac{\delta H_{\mu\nu}^{02}(\omega)-F_{\mu\nu}^{02}(\omega)}{E_\mu + E_\nu+\omega} \,.
\end{aligned}
\label{fam-qrpa}
\end{equation}
Since induced matrices $\delta H_{\mu\nu}^{20}$ and $\delta H_{\mu\nu}^{02}$ depend on $X_{\mu\nu}$ and $Y_{\mu\nu}$ amplitudes, 
one has to employ an iterative, self--consistent scheme to solve the FAM--QRPA equations. 

Formally, the FAM--QRPA system (\ref{fam-qrpa}) is equivalent to the linear response theory, 
\begin{align}
R(\omega)^{-1} 
 \begin{pmatrix}
  X(\omega) \\
  Y(\omega) 
 \end{pmatrix}
 = -
 \begin{pmatrix}
  F^{20} \\
  F^{02} 
 \end{pmatrix},
 \label{linearResponse}
\end{align}
where the response function $R(\omega)$ can be expressed as
\begin{align}
R(\omega)^{-1} =\left[
\begin{pmatrix}
  A      & B     \\
  B^\ast & A^\ast 
\end{pmatrix}
 -\omega
\begin{pmatrix}
  1 &  0     \\
  0 & -1 
\end{pmatrix} 
\right], 
\end{align}
where $A$ and $B$ are the well--known QRPA matrices. If the external field is set to zero, the linear response equation  (\ref{linearResponse}) will transform to the standard matrix QRPA equation
\begin{align}
\begin{pmatrix}
  A      & B     \\
  B^\ast & A^\ast 
\end{pmatrix}
\begin{pmatrix}
  X(\omega) \\
  Y(\omega) 
 \end{pmatrix}
 =\omega
 \begin{pmatrix}
  X(\omega) \\
  -Y(\omega) 
 \end{pmatrix}.
\end{align}
Since matrices $A$ and $B$ have typically very large dimensions in the deformed case, solving the above equation is computationally rather demanding. 
The essential asset of the FAM--QRPA lies in the fact that instead of a very time--consuming construction and
diagonalization of the large QRPA matrix within the standard MQRPA procedure, one calculates the FAM--QRPA amplitudes 
iteratively with respect to the response of the self--consistent Hamiltonian, 
i.e. induced fields $\delta H^{20}$ and $\delta H^{02}$, to the external field. 
This significantly reduces the computational cost in comparison to the MQRPA.  

The transition strength function of the operator $\hat F$ at the frequency $\omega$ is defined as
\begin{align}
\frac{d B(\hat F ;\omega)}{d \omega} = -\frac{1}{\pi} \text{Im}\, S(\hat F ;\omega),
\end{align}
where the FAM--QRPA strength function $S(\hat F ;\omega)$ can be expressed as
\begin{align}
S(\hat F;\omega) &= \phantom{-}\sum_{\mu<\nu} \left[F_{\mu\nu}^{20\ast} X_{\mu\nu}(\omega)+F_{\mu\nu}^{02\ast}Y_{\mu\nu}(\omega)\right] \,.
\label{eq:Str}
\end{align}
In order to obtain a finite value for the FAM--QRPA transition strength, a small imaginary part,
$\omega \rightarrow \omega_\gamma = \omega + i\gamma$, is added to the excitation energy. 
This leads to a Lorentzian smearing of transition strength function with a width of $\Gamma = 2\gamma$. 
However, in order to access a Nambu--Goldstone mode, one needs to evaluate strength function (\ref{eq:Str}) at the vanishing frequency 
$\omega_\gamma=\omega=0$. We will discuss this relation in the next subsection.    

\subsection{Collective Thouless--Valatin inertia and Nambu--Goldstone modes}

\begin{figure*}
\centering
    \includegraphics[width=\linewidth]{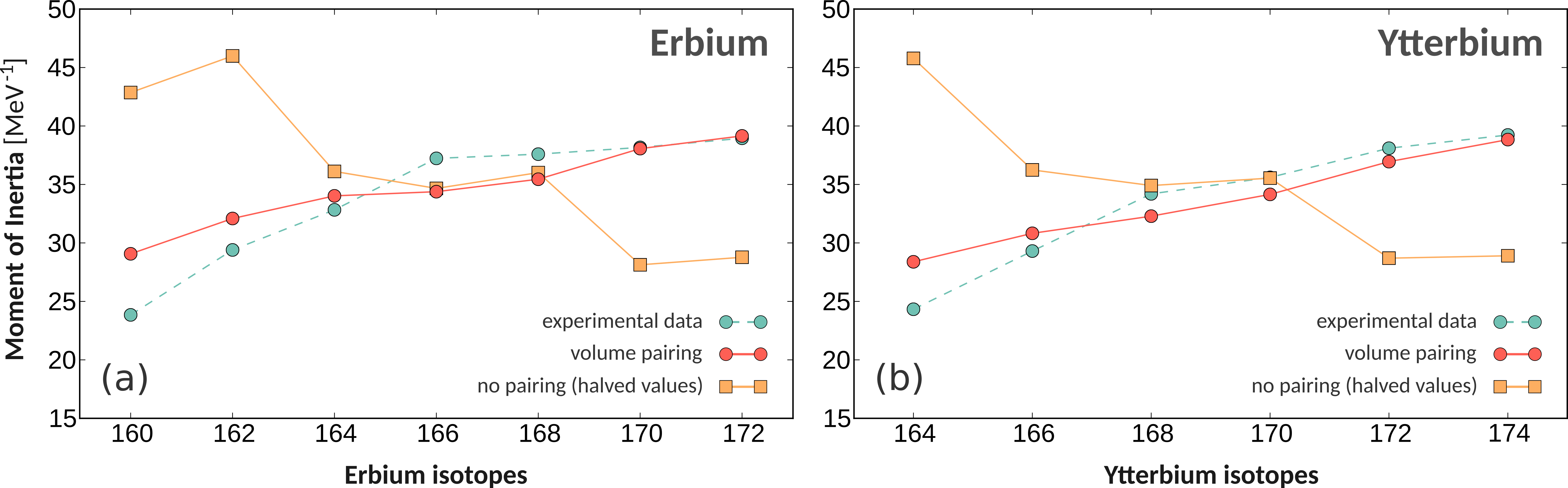}
\caption{Thouless--Valatin rotational moment of inertia in \textit{erbium} (a) and \textit{ytterbium} (b) isotopic chains. Full lines indicate the volume pairing and vanishing pairing options; in the latter case, the values of the TV inertia are divided by $2$ in order to make a comparison with other results. The dashed lines show the experimental data from the rotational bands ($2^+$ states). For erbium isotopes, the volume pairing was adjusted to the experimental proton and neutron pairing gaps of $^{166}$Er, while for ytterbium isotopes, the pairing gaps of $^{168}$Yb were chosen.}
\label{fig:ErYbInertia}
\end{figure*}

When some form of a mean--field approximation is introduced, a spontaneous symmetry breaking phenomenon that contains information about correlations to the one--body approximation can occur. Continuous symmetries conserved by the exact many--body Hamiltonian, which may be broken spontaneously, are the \textit{translational}, \textit{rotational} and \textit{particle number} gauge symmetry~\cite{Ring80}. In addition, the \textit{isospin} symmetry is broken explicitly by the presence of the Coulomb interaction
and spontaneously due to the mean--field~\cite{Sat11}.

As a result, spontaneous breaking of any of the above continuous symmetries leads to the appearance of a new zero--energy mode that restores the given symmetry. Such zero--energy restoration modes are called \textit{Nambu--Goldstone} (NG) modes and are a consequence of the symmetry--broken mean--field and represent a special case of collective motion \cite{Nambu60,Goldstone61,Ring80}. They are also associated with the infinitesimal transformation of the frame of reference and since they do not represent real physical excitations in the intrinsic frame, they are often called \textit{spurious} modes.  

It is well known that breaking of the translational symmetry introduces the \textit{center of mass} NG mode, rotational symmetry the \textit{rotational} NG mode and particle number gauge symmetry the \textit{pairing--rotational} NG mode. In the intrinsic frame, each NG mode can be associated with a collective inertia that has experimental correspondence to an actual spectroscopy measurement in the laboratory frame. In the QRPA framework, such collective inertia coupled to the NG mode is called the \textit{Thouless--Valatin} inertia \cite{Thouless62, Hinohara15}. For the translational (center of mass) NG mode the TV inertia is simply the total mass of the nucleus and is easily obtainable since the coordinate and momentum QRPA phonon operators, needed for the estimation, are known in advance. In the case of the (pairing--)rotational NG mode the TV inertia represents the nuclear (pairing--)rotational moment of inertia of the nucleus.

With the exception of the center of mass NG mode, in order to obtain the TV inertia one has to fully solve the QRPA equations. Although all necessary expressions are known in the standard matrix QRPA formalism, the full evaluation of the QRPA $A$ and $B$ matrices is computationally heavy, mostly due to their large dimensions. Other equivalent approaches have been therefore employed to obtain this quantity. The most important ones are the perturbation expansion procedure in the adiabatic time--dependent HFB theory, the cranked mean--field calculations within the HFB theory and most recently, the FAM--QRPA approach developed for the nuclear density functional theory. 

The FAM--QRPA has been already successfully formulated for calculating the symmetry restoring NG modes of the translational and particle number gauge symmetries~\cite{Hinohara15,Hin16}. It has been shown that the TV inertia calculated in this way provides crucial information about the ground state correlations to the relevant broken symmetry and that the FAM--QRPA represents a very precise and effective method to study this quantity. Our present work continues in this direction and extends the FAM--QRPA also to the case of the rotational NG mode and related rotational TV moment of inertia.   

To obtain information about the given NG mode one can use a relevant one--body operator as the external field and evaluate the value of the strength function at the zero frequency. In the case of the translational symmetry, which introduces the center of mass NG mode, both the coordinate $\hat Q$ and momentum operators $\hat P$ are known and the TV inertia is simply the total mass of the studied nucleus, $M_{\text{NG}}=A m$, where $A$ is the atomic mass number and $m$ mass of the nucleon. This is the only example where it is not necessary to solve QRPA equations to obtain the TV inertia.

The aim of this work is to study the rotational Thouless--Valatin moment of inertia of the rotational NG mode, where the one--body operator of the interest is the total angular momentum. Depending on the symmetries chosen, a specific component $\hat J_i$ of the angular momentum operator has to be selected. 
Such a component, when used as the external perturbation, leads to the rotational TV inertia that can be extracted from the 
strength function at zero frequency as
\begin{align}
S(\hat J_i;0)_{\text{NG}} &= \sum_{\mu<\nu} \left[(J^{20\ast}_i)_{\mu\nu} X_{\mu\nu}(0)+(J^{02\ast}_i)_{\mu\nu}Y_{\mu\nu}(0)\right]
\nonumber\\
&= -M_{\text{NG}}^{J_i} \,.
\end{align}
Here, $M_{\text{NG}}^{J_i}$ is the TV moment of inertia along the $i$-axis.
 
\section{NUMERICAL RESULTS}

\begin{figure*}
\centering
    \includegraphics[width=\linewidth]{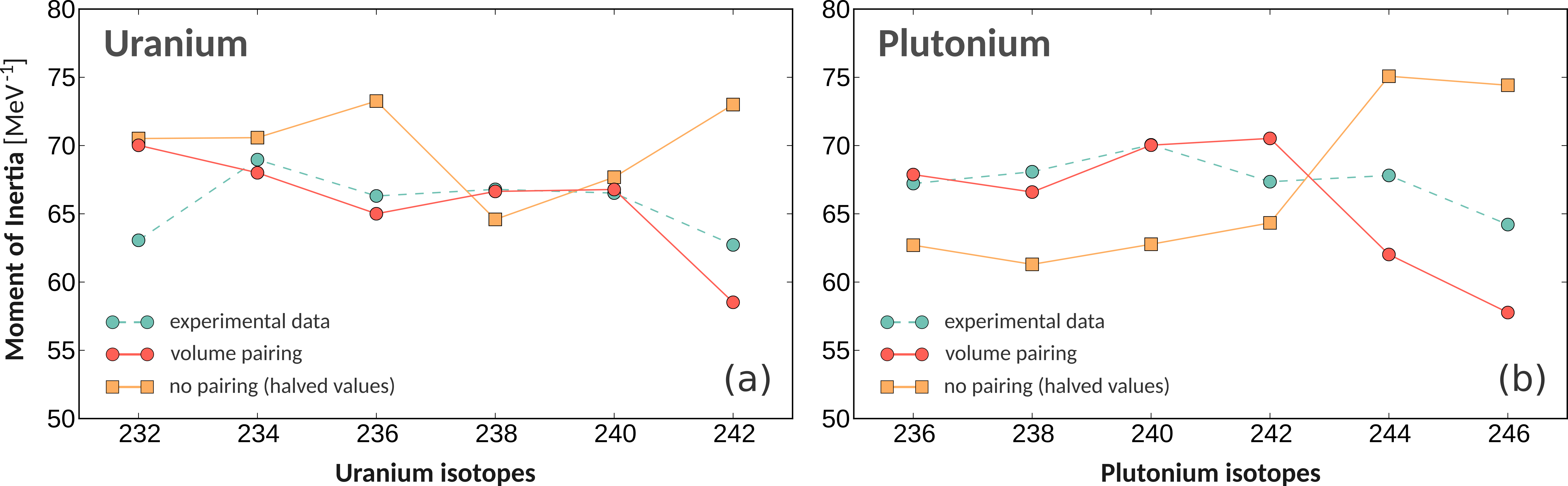}
\caption{The same as Fig.~\ref{fig:ErYbInertia} but for \textit{uranium} (a) and \textit{plutonium} (b) isotopic chains. 
For uranium isotopes, the volume pairing was adjusted to the experimental proton and neutron pairing gaps of $^{238}$U, 
while for plutonium isotopes, the pairing gaps of $^{240}$Pu were chosen.}
\label{UPuInertia}
\end{figure*}

Our calculations were carried out using the FAM--QRPA code on top of the HFB code, specifically the computer program \textsf{HFBTHO} \cite{Stoitsov13}, 
which is an HFB solver providing axially symmetric solutions using the harmonic oscillator basis with the Skyrme energy density functional. 

For the HFB calculations, and subsequent FAM-QRPA calculations, we have employed the Skyrme SkM* parameter set~\cite{skms} at the
particle--hole channel. This parameter set is know to be stable to the linear response in infinite nuclear matter~\cite{Pas12}.
The standard definition of the Skyrme EDF and associated densities can be found e.g. in Ref.~\cite{Bender03}.
For the pairing part, the particle--particle channel interaction was taken to be a simple contact interaction without a density dependence, 
yielding the pairing energy density commonly called the \textit{volume pairing}. 
In all studied cases, this type of pairing provided generally better reproduction of the data over the density dependent mixed pairing.
A specific nucleus in a given isotopic chain was chosen to adjust the pairing strengths separately for neutrons and protons in order 
to reproduce the empirical pairing gaps. Due to the used zero--range pairing interaction, it was necessary to introduce a pairing window to prevent divergent energy in the pairing channel.
In this work, the quasiparticle cut--off energy was set to 60\,MeV in all considered cases. 
The deformed HFB ground state obtained through this setup breaks the translational, rotational and particle number symmetries.

One of the main assets of the FAM--QRPA formalism is that no additional truncation or cut--offs of the two--quasiparticle space are imposed. This guarantees the full self--consistency between the QRPA solution and the HFB ground state. Even though the underlying HFB calculations have
conserved the time--reversal symmetry, the full time--odd part of the EDF was used at the FAM--QRPA level.
The FAM module implementation followed that of Ref.~\cite{Kortelainen15} and the FAM equations were solved in the presence of the simplex--$y$ symmetry~\cite{Dobaczewski00}.

To ensure a satisfactory convergence, we have selected 20 major oscillator shells for the HO basis in all our calculations and confirmed that no relevant improvement could be achieved from employing more shells. The FAM--QRPA iterative solution was obtained using the very efficient \textit{modified Broyden method} \cite{Broyden65, Johnson88, Baran08} that provides stable and fast convergence, especially when multiprocessor tasks are considered.  
Values of the spatial integrals were numerically approximated using the Gauss--Hermite (mesh--point number set to $N_{\text{GH}}=80$) and Gauss--Laguerre (mesh--point number set to $N_{\text{GL}}=40$) quadratures. 

In this work, we focus on two areas of the nuclear chart, the heavy rare--earth isotopes and heavy actinide isotopes. Both regions offer plenty of information about collective properties and rotational characteristics of axially deformed nuclei. 

\subsection{Rotational Thouless--Valatin inertia}

To obtain the TV collective inertia we need to calculate the strength function of the symmetry restoring NG mode and take the value at zero energy (zero frequency). Since we are interested in the rotational TV inertia under the assumption of the axial symmetry and simplex--$y$ symmetry at the HFB level, we focus on the response to the $y$--component of the total angular momentum operator, which is a combination of the $y$--components of the orbital and spin angular momentum operators $\hat J_y = \hat L_y + \hat S_y$. 
In order to investigate the role of the pairing with respect to the TV inertia, we compare the calculations where pairing correlations are taken into account with those calculations of vanishing pairing.

First, we have studied the rare--earth region of the nuclear chart, specifically the \textit{erbium} (even--even nuclides from $^{160}$Er to $^{172}$Er) and \textit{ytterbium} (even--even nuclides from $^{164}$Yb to $^{174}$Yb) isotopic chains. In the case of erbiums, the pairing strength parameters ($V_0^p=-211.20$\,MeV\,fm$^{3}$ and $V_0^n=-178.83$\,MeV\,fm$^{3}$) were adjusted separately to $^{166}$Er proton and neutron pairing gaps of $1.20$\,MeV and $1.02$\,MeV, respectively, 
obtained from the three--point mass difference formula~\cite{Ber09}.
For ytterbiums, we adjusted the pairing strengths ($V_0^p=-212.40$\,MeV\,fm$^{3}$ and $V_0^n=-180.60$\,MeV\,fm$^{3}$) to $^{168}$Yb proton and neutron pairing gaps of $1.26$\,MeV and $1.09$\,MeV, respectively. 

Consequently, we have calculated the TV inertia using the same input parameters, but with the vanishing pairing option chosen (pairing strengths were selected to be negligibly small to induce a collapse of the pairing) and compared both sets of results to the experimental moments of inertia extracted from the $2^+$ state 
energy of the ground--state rotational bands of studied isotopes \cite{ENSDF}. This comparison can be found in Fig.~\ref{fig:ErYbInertia}. 

\begin{figure*}
\centering
    \includegraphics[width=\linewidth]{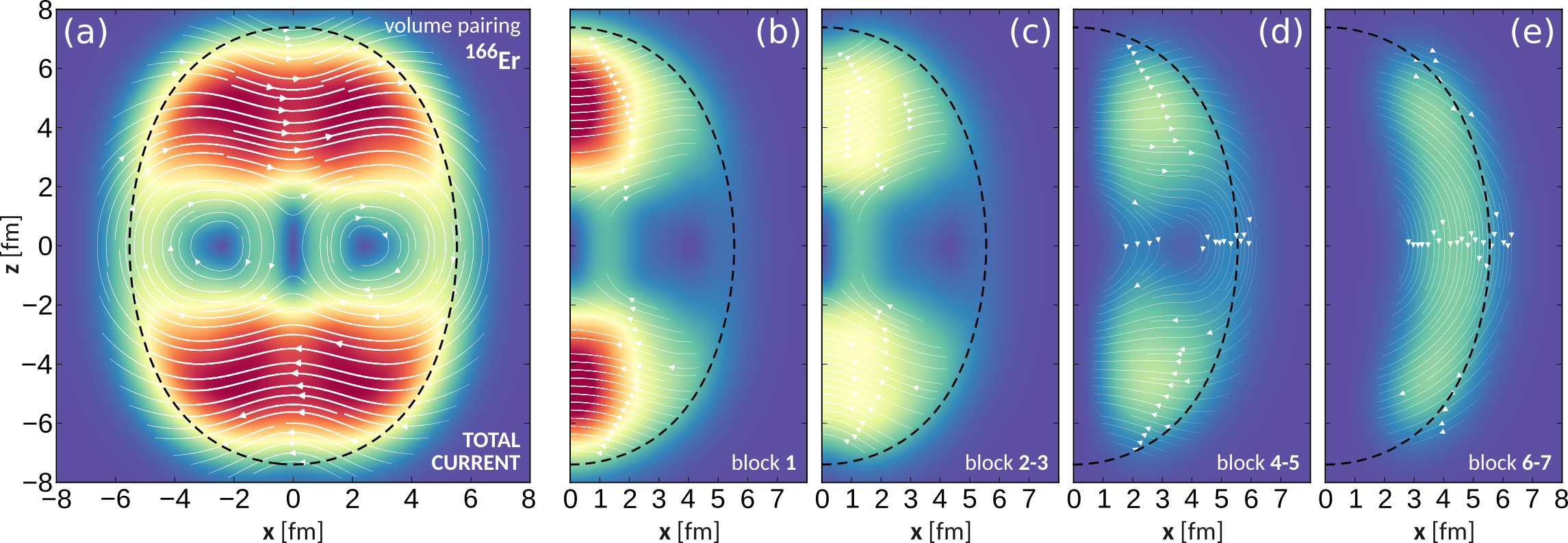}
\caption{The flow (lines with arrows) of the induced isoscalar current density $\vec{j}_0$ and its magnitude (color contours and thickness of the flow line) for $^{166}$Er with volume pairing setup are shown in the $x$--$z$ plane. The full cross--section (a) shows the total current density, while the half--plots (b)--(e) indicate the partial induced currents as given by the specific QRPA blocks. The black dashed contour line indicates the nuclear surface at matter density of $\rho_0=0.08\,{\rm fm}^{-3}$. See the text for further details.}
\label{flow1}
\end{figure*}

When the volume pairing is considered, the reproduction of the experimental moment of inertia by the TV rotational inertia from FAM--QRPA is good for both erbium and ytterbium isotopic chains. It can be easily demonstrated that by adjusting the pairing strengths of all isotopes to the pairing gaps of a different nucleus from the chain, the values of the TV inertia shift up or down, basically retaining the overall pattern. Therefore, only partial improvements can be achieved by changing the pairing strengths from case to case with no significant impact on the descriptive or predictive capability.

When the nuclear pairing is not taken into account, the calculated TV inertia values are markedly higher (in Fig.~\ref{fig:ErYbInertia}, values without pairing are divided by 2 for the sake of comparison). We can conclude that the pairing correlations have crucial effects on the rotational collective motion and the TV inertia can be several times higher in contrast to the superfluid calculations.      

In the next step, we focused on the actinide region of the nuclear chart, where we picked the \textit{uranium} (even--even nuclides from $^{232}$U to $^{242}$U) and \textit{plutonium} (even--even nuclides from $^{236}$Pu to $^{246}$Pu) isotopic chains. Here we again studied the effects of the pairing correlations on the rotational TV inertia and compared cases with and without pairing. The pairing strengths ($V_0^p=-216.18$\,MeV\,fm$^{3}$ and $V_0^n=-170.95$\,MeV\,fm$^{3}$) for uraniums were adjusted to $^{238}$U proton and neutron pairing gaps of $1.11$\,MeV and $0.67$\,MeV, and pairing strengths ($V_0^p=-213.20$\,MeV\,fm$^{3}$ and $V_0^n=-170.80$\,MeV\,fm$^{3}$) for plutoniums to $^{240}$Pu proton and neutron gaps of $0.99$\,MeV and $0.64$\,MeV, respectively. Results are shown in Fig.~\ref{UPuInertia}.  

Once again, we obtain rather good agreement between the experiment and the rotational inertia obtained by the FAM--QRPA approach. One can see especially precise, although presumably coincidental reproduction in the case of $^{238}$U and $^{240}$Pu, which were the isotopes chosen for the pairing strength adjustments. The TV inertia values for collapsed pairing (values divided by 2 in the plots) are again notably higher, indicating the importance of the pairing.     
  
In order to test the precision and validity of our FAM--QRPA approach, we carried out several independent calculations using a 
different HFB solver (the \textsf{HFODD} code~\cite{hfodd}) and extracted the moment of inertia by using the cranking formalism. Since the HFB cranking calculations are rather time--consuming, we chose 12, 14, and 16 major oscillator shells for our setup and erbium isotope $^{166}$Er as our test nucleus. These calculations were compared to the 12--, 14--, and 16--shell FAM--QRPA results. In all cases the same SkM$^\ast$ parametrization was used together with identical pairing strengths.

\begin{figure*}
\centering
    \includegraphics[width=\linewidth]{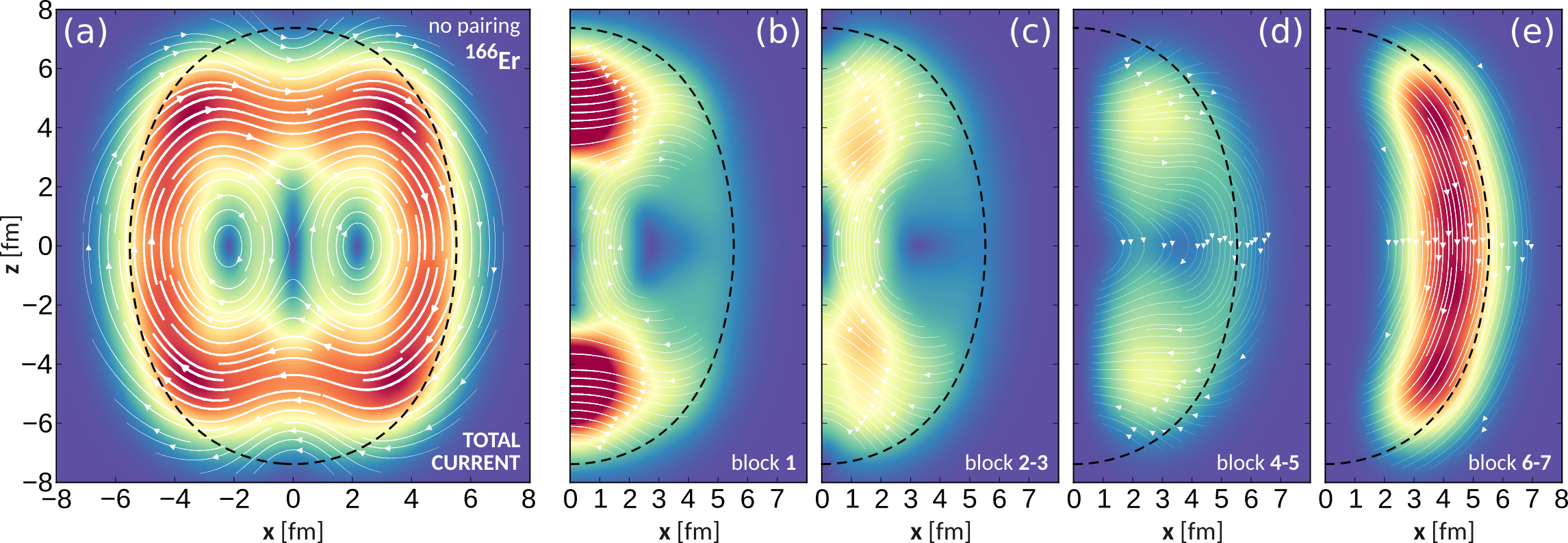}
\caption{The same as Fig.~\ref{flow1}, but with vanishing pairing.}
\label{flow2}
\end{figure*}

The TV moments of inertia from the cranking formula were obtained by extrapolation to the zero cranking frequency based on the behavior of the inertia for non--zero frequencies. We have obtained an excellent agreement in all studied cases, for example, in the case of 16--shell calculation, the FAM--QRPA TV inertia yields a value of $34.511$\,MeV$^{-1}$, while the cranking moment of inertia equals $34.533$\,MeV$^{-1}$.
See supplementary material for more detailed results.

This excellent correspondence between two calculations underlines the efficiency and usefulness of the FAM--QRPA formalism in calculating the rotational TV inertia and we can confirm that our approach is very convenient and computationally significantly faster than standard cranking HFB calculations.

\subsection{Induced current density}

In this section, we investigate the collective rotational behavior originating from the total angular momentum operator more closely. 
As discussed previously, the NG mode, related to the spontaneous symmetry breaking of the rotational symmetry, leads to the TV moment of inertia connected to the strength function of the angular momentum operator at zero frequency. Application of this operator as an external perturbation gives rise to the induced densities and currents of various types. Here we want to have a look on the induced \textit{isoscalar current density}, $\vec {j}_0 \equiv \vec {j}_{\rm n}+\vec {j}_{\rm p}$, which can give us some hints about the movement of protons and neutrons in the studied isotopes. We have selected the $^{166}$Er nucleus as an example and studied the cases with and without pairing present. 

In Fig.~\ref{flow1}, one can see the $x$--$z$ plane cross--section of the nucleus (panel (a)), where the total magnitude (color contours and line thickness) 
along with the directions (lines with arrows) of the total induced current density are shown. 
The right side of Fig.~\ref{flow1} (panels (b)--(e)) shows the partial contributions coming from the first seven 
QRPA blocks with values normalized to the same maximum value to make comparisons possible. 
These blocks give the largest contribution to the total TV moment of inertia. When two block numbers 
are indicated (e.g. block $2$--$3$), there is only one plot shown since the contribution is identical from both QRPA blocks. 
Specifically, the block $1$ corresponds to quasiparticle transitions between $\abs{\Omega}=\tfrac{1}{2}$ quasiparticle states. 
Blocks $2$ and $3$ correspond to transitions between $\abs{\Omega}=\tfrac{1}{2}$ and $\abs{\Omega}=\tfrac{3}{2}$ quasiparticle 
states, blocks $4$ and $5$ correspond to transitions between $\abs{\Omega}=\tfrac{3}{2}$ and $\abs{\Omega}=\tfrac{5}{2}$ states, and so on.
The $\Omega$ quantum number denotes the total angular momentum projection of the quasiparticle state along the $z$--axis.
In this way, one can understand the importance of quasiparticle states of a given $\Omega$ and their impact on the total induced current. This decomposition is rather useful when the comparison between the cases with and without pairing is made. In order to examine this, we have calculated the induced current density and matter density under the vanishing pairing setup as well. They can be found in Fig.~\ref{flow2}. 

We already know that the presence of the pairing correlations significantly lowers the value of the rotational TV inertia and thus one would expect to observe this effect also in the induced current density. Comparing the full plots in Figs.~\ref{flow1} and~\ref{flow2}, we clearly see that the flow in the case with pairing resembles the irrotational flow behavior, whereas the case without pairing is noticeably closer to the rigid body rotation. 
Similar kind of flow patters also emerge with the rotation of trapped atomic gases~\cite{Urban2003}, where the temperature 
has a critical impact on the moment of inertia and the current flow.

In order to track down the source of this effect, we analyzed the contributions from the QRPA blocks separately. Looking at the 
panels (b)--(e) of Figs.~\ref{flow1} and~\ref{flow2}, we see that although the first couple of QRPA blocks are basically similar, a strong effect can be seen in the blocks 6 and 7 in the case without pairing. Here, we observe the contribution that explains the major difference discussed above. The vanishing pairing enhances the rotational component, which shifts the irrotational flow structure towards the rigid body flow pattern. This shows that the role of high--$\Omega$ states becomes more prominent with vanishing pairing 
when compared to the superfluid case.
Similar kind of behavior occurs in the rotational flow of $^{240}$Pu and also in the flow of the oblate minimum of $^{166}$Er, see the supplementary material.
We did not see any large qualitative differences between the proton and neutron flow apart from the fact that the neutron flow was stronger in general due to higher abundance of neutrons.

\begin{figure}
\centering
    \includegraphics[width=\linewidth]{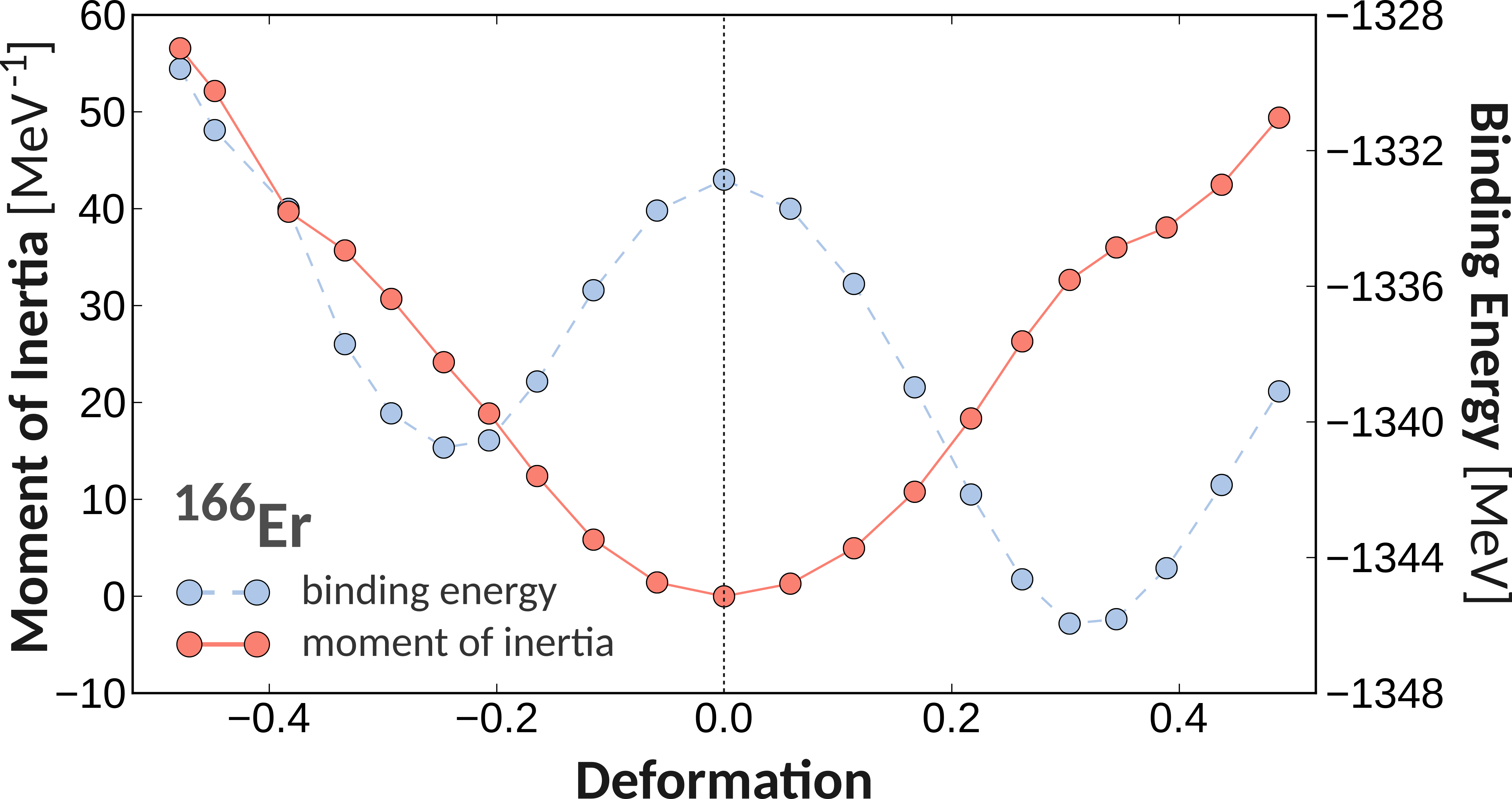}
    \caption{Calculated Thouless--Valatin rotational moment of inertia and the HFB binding energy of $^{166}$Er, both as function of the deformation parameter.}
\label{Erbium}
\end{figure}

\subsection{Deformation dependence of TV inertia}

Rare--earth isotopes described in the present work are deformed nuclei with a moderate prolate deformation. An example of such prolate ground state 
can be easily seen from the plot of the HFB binding energy versus deformation dependence of $^{166}$Er, as shown in Fig.~\ref{Erbium}.
Here we can see that in addition to the prolate minimum, there exists also a higher--lying oblate minimum, therefore, it is interesting to examine the TV rotational inertia dependence on the deformation as well. To obtain the results for $^{166}$Er seen in Fig.~\ref{Erbium}, 
we carried out constrained HFB calculations and scanned deformation from $-0.5$ to $+0.5$. For each deformation point we calculated the TV inertia in the FAM--QRPA framework. As expected, the TV inertia vanishes in the spherical case, when the deformation is set to zero, however, once the deformation is present, it increases smoothly with the increasing deformation in both, prolate and oblate cases.

One can observe small dips in the moment of inertia, which appear when the deformation is increased farther away from both minima. These dips are a direct consequence of the behaviour of pairing gaps and pairing energy with changing deformation. They are linked to the local increase of the pairing gaps and related increase of the pairing energy in the region we are focusing on. Such a feature supports the general picture and conclusions about the role of the superfluidity in the properties of the rotational TV inertia.  

These calculations are of great interest, since the rotational moment of inertia is one of the vital inputs for the microscopic collective Hamiltonian model \cite{Rohozinski12,Matsuyanagi16} that requires such local QRPA results. Our work proves the feasibility of the FAM--QRPA for calculating collective mass parameters of this type. The FAM--QRPA framework is thus an excellent candidate for furnishing necessary quantities for this kind of models.

\section{CONCLUSION}

In the present work, we have studied the small--amplitude collective nuclear motion related to the rotational moment of inertia, which is a consequence of a response to the total angular momentum operator acting in the role of the external field. We have successfully extended the FAM--QRPA approach to the case of the rotational NG mode and demonstrated the numerical efficiency in calculating important nuclear properties. 
This work also paves a way for the removal of the spurious component from the transition strength function, appearing with $K^{\pi}=1^+$ type
of operators when deformation is present.

Our calculations covered several axially deformed nuclei from the rare--earth and heavy actinide regions. The analysis of the rotational Thouless--Valatin moment of inertia has shown the importance of the pairing correlations in the description of the rotational attributes of superfluid nuclei. When the pairing strength was adjusted to the experimental pairing gaps, the rotational moment of inertia for most of the isotopes was reproduced very well. 
The high--$\Omega$ quasiparticle states were found to have a major role in increasing the moment of inertia value in the case with no pairing considered.

Lastly, we demonstrated that FAM--QRPA can be used to extract the rotational moment of inertia within the constrained HFB calculations.
Such kind of local QRPA calculations are essential when constructing a microscopic collective Hamiltonian model.
As a future avenue, we therefore plan to employ the FAM--QRPA approach to provide such microscopic input for a collective Hamiltonian, 
with a purpose to apply it to the large amplitude collective motion in nuclei.

\section*{ACKNOWLEDGMENTS}
We thank Jacek Dobaczewski, Karim Bennaceur and Nobuo Hinohara for valuable discussions and helpful observations.
This work has been supported by the University of Jyv\" askyl\" a and Academy of Finland under the Centre of
Excellence Program 2012--2017 (Nuclear and Accelerator--Based Physics Program at JYFL) and FIDIPRO programme.
We acknowledge CSC-IT Center for Science Ltd., Finland, for the allocation of computational resources.

\end{document}